\documentclass[aps,prl,twocolumn,superscriptaddress,showpacs]{revtex4}
\usepackage{amssymb}
\usepackage{txfonts}
\usepackage{tipa}

\usepackage{graphicx}
\usepackage{dcolumn}
\usepackage{bm}

\begin{document}


\title{Magnetic structure of EuFe$_2$As$_2$ determined by single crystal neutron diffraction}

\author{Y. Xiao}
\email[y.xiao@fz-juelich.de]{}
\affiliation{Institut fuer Festkoerperforschung, Forschungszentrum
Juelich, D-52425 Juelich, Germany}

\author{Y. Su}
\affiliation{Juelich Centre for Neutron Science, IFF,
Forschunszentrum Juelich, Outstation at FRM II, Lichtenbergstrasse
1, D-85747 Garching, Germany}

\author{M. Meven}
\affiliation{FRM II, Technische Universitaet Muenchen,
Lichtenbergstrasse 1, D-85747 Garching, Germany}

\author{R. Mittal}
\affiliation{Juelich Centre for Neutron Science, IFF,
Forschunszentrum Juelich, Outstation at FRM II, Lichtenbergstrasse
1, D-85747 Garching, Germany}

\affiliation{Solid State Physics Division, Bhabha Atomic Research
Centre, Trombay, Mumbai 400 085, India}

\author{C.M.N. Kumar}
\affiliation{Institut fuer Festkoerperforschung, Forschungszentrum
Juelich, D-52425 Juelich, Germany}

\author{T. Chatterji}
\affiliation{Juelich Centre for Neutron Science, Forschungszentrum
Juelich, Outstation at Institut Laue-Langevin, BP 156, 38042
Grenoble Cedex 9, France}

\author{S. Price}
\affiliation{Juelich Centre for Neutron Science, IFF,
Forschunszentrum Juelich, Outstation at FRM II, Lichtenbergstrasse
1, D-85747 Garching, Germany}

\author{J. Persson}
\affiliation{Institut fuer Festkoerperforschung, Forschungszentrum
Juelich, D-52425 Juelich, Germany}

\author{N. Kumar}
\affiliation{Department of Condensed Matter Physics and Material
Sciences, Tata Institute of Fundamental Research, Homi Bhabha Road,
Colaba, Mumbai 400 005, India}

\author{S. K. Dhar}
\affiliation{Department of Condensed Matter Physics and Material
Sciences, Tata Institute of Fundamental Research, Homi Bhabha Road,
Colaba, Mumbai 400 005, India}

\author{A. Thamizhavel}
\affiliation{Department of Condensed Matter Physics and Material
Sciences, Tata Institute of Fundamental Research, Homi Bhabha Road,
Colaba, Mumbai 400 005, India}

\author{Th. Brueckel}
\affiliation{Institut fuer Festkoerperforschung, Forschungszentrum
Juelich, D-52425 Juelich, Germany} \affiliation{Juelich Centre for
Neutron Science, IFF, Forschunszentrum Juelich, Outstation at FRM
II, Lichtenbergstrasse 1, D-85747 Garching, Germany}
\affiliation{Juelich Centre for Neutron Science, Forschungszentrum
Juelich, Outstation at Institut Laue-Langevin, BP 156, 38042
Grenoble Cedex 9, France}

\date{\today}

\begin{abstract}

Among various parent compounds of iron pnictide superconductors,
EuFe$_2$As$_2$ stands out due to the presence of both spin density
wave of Fe and antiferromagnetic ordering (AFM) of the localized
Eu$^{2+}$ moment. Single crystal neutron diffraction studies have
been carried out to determine the magnetic structure of this
compound and to investigate the coupling of two magnetic
sublattices. Long range AFM ordering of Fe and Eu spins was observed
below 190 K and 19 K, respectively. The ordering of Fe$^{2+}$
moments is associated with the wave vector \textbf{k} = (1,0,1) and
it takes place at the same temperature as the tetragonal to
orthorhombic structural phase transition, which indicates the strong
coupling between structural and magnetic components. The ordering of
Eu moment is associated with the wave vector \textbf{k} = (0,0,1).
While both Fe and Eu spins are aligned along the long \emph{a} axis
as experimentally determined, our studies suggest a weak coupling
between the Fe and Eu magnetism.

\end{abstract}

\maketitle

\section{\label{sec:level1}I. Introduction}

The recent discovery of pnictide superconductors has drawn extensive
attention because it provides a new opportunity to investigate the
mechanism of superconductivity
\cite{Kamihara,Takahashi,Chen1,Matsuishi1,Rotter1}. Most of the
research on pnictide superconductors has focused on
\emph{R}FeAs(O$_x$F$_{1-x}$)(with \emph{R} = La, Nd and Sm) and
\emph{A}Fe$_2$As$_2$ (with \emph{A} = Ba, Ca and Sr), the so called
'1111' and '112' families. These two families are closely related
since both of them adopt a layered structure with a single FeAs
layer in the unit cell of '1111' and two such layers in the unit
cell of '122'. The superconducting state can be achieved either by
electron or hole doping of the parent compounds
\cite{Wen,Ren,Matsuishi2}. Till now, the highest \emph{T}$_c$
attained is 57.4 K in the electron doped Ca$_{0.4}$Na$_{0.6}$FeAsF
'1111' compound \cite{Cheng}, while for '122' family the highest
\emph{T}$_c$ of 38 K is reached in the hole doped
Ba$_{0.6}$K$_{0.4}$Fe$_2$As$_2$ \cite{Rotter2}. Considering that the
electronic states near the Fermi surface are dominated by
contributions from Fe and As, it is believed that the FeAs layers
are responsible for superconductivity in these compounds.

Recent neutron diffraction experiments reveal that the formation of
the spin density wave (SDW), originating from the long range
antiferromagnetic (AFM) order of the Fe moments at low temperature,
seems to be a common feature for all the iron pnictide parent
compounds \cite{Cruz,Huang,Su1,Goldman,Xiao}. The onset of the AFM
order is also accompanied by the Tetragonal-Orthorhombic (T-O)
structural phase transition in the '122' family and preceded by the
T-O phase transition for the '1111' family. Phase diagrams of some
iron pnictides clearly show that the magnetic order is suppressed
with appropriate charge carrier doping of parent compound.
Concomitantly, superconductivity emerges and reaches a high
\emph{T}$_c$ at optimal doping~\cite{Zhao1}, thus exhibiting
features similar to high \emph{T}$_c$ cuprates \cite{Bednorz}. It is
generally believed  that the superconductivity in iron pnictides is
unlikely due to simple electron-phonon coupling, as demonstrated
from extensive studies of phonon dynamics \cite{Ranjan1,Ranjan2}.
Magnetism seems to play a crucial role in the appearance of
superconductivity and AFM spin fluctuations have thus been suggested
to be a possible paring mechanism. Strong evidence on the presence
of resonant spin excitation in the superconducting phase has indeed
been obtained from recent inelastic neutron scattering experiments
on several optimally doped '122' superconductors
\cite{Christianson,Lumsden,Chi,Su2}.

EuFe$_2$As$_2$ is a peculiar member of iron arsenide
\emph{A}Fe$_2$As$_2$ family since the \emph{A} site is occupied by
Eu$^{2+}$, which is an \emph{S}-state (orbital angular momentum
\emph{L} = 0) rare-earch ion possessing a 4\emph{f}$^7$ structure
with the electron spin \emph{S} = 7/2. The theoretical effective
magnetic moment of Eu$^{2+}$ ion is 7.94 $\mu$$_B$. As revealed by
M\"{o}ssbauer and magnetic susceptibility measurement on single
crystals, the Eu$^{2+}$ moments order antiferromagneticlly below
\emph{T}$_N$ $\sim$ 20 K \cite{Raffius,Ren2}. It is also reported
that the moment of Eu$^{2+}$ can be realigned ferromagneticly by
applying a magnetic field \cite{Jiang,Wu}. Besides,
superconductivity can also be achieved by replacing Eu by alkali
metals, \emph{e.g.} the \emph{T}$_c$ is observed to be 31 K and 34.7
K for Eu$_{0.5}$K$_{0.5}$Fe$_2$As$_2$ \cite{Jeevan1} and
Eu$_{0.7}$Na$_{0.3}$Fe$_2$As$_2$ \cite{Qi}, respectively. Unlike
BaFe$_2$As$_2$, in which the superconductivity emerges with the Ni
substitution of Fe, the SDW is suppressed in
EuFe$_{2-x}$Ni$_x$As$_2$ without the emergence of
superconductivity~\cite{Ren3}. Furthermore, the magnetic ordering of
Eu$^{2+}$ moments evolves from AFM to ferromagnetic at higher levels
of Ni doping.

Since magnetism and superconductivity appears to be intimately
related in iron pnictides, it is therefore equally  important to
understand the magnetic properties especially for the compounds that
contain the magnetic lanthanide ions. The investigation of the
interplay between the lanthanide and iron magnetism may also be
crucial for a deeper understanding of the magnetic and electronic
properties of iron pnictides. For EuFe$_2$As$_2$, the magnetic
ordering and the details of magnetic structure have not been
clarified so far  via single-crystal neutron diffraction due to the
extremely large neutron absorption cross-section of Eu. Here we
report neutron diffraction studies on a high-quality EuFe$_2$As$_2$
single crystal using the hot neutron source. It has been observed
that both the Fe$^{2+}$ and Eu$^{2+}$ moments are ordered
antiferromagnetically below 190 K and 19 K, respectively. A unique
propagation vector \textbf{k} = (1,0,1) is determined for the Fe
magnetic sublattice with the moment aligned along the \emph{a} axis.
Furthermore, the magnetic propagation vector is determined to be
\textbf{k} = (0,0,1) for the Eu$^{2+}$ moment, which is also aligned
along the \emph{a} axis. The coupling between the Fe and Eu magnetic
sublattices has been found to be rather weak. The determination of
the magnetic structure of EuFe$_2$As$_2$ would pave the way for
further investigations of EuFe$_2$As$_2$ under high pressure and
strong magnetic fields.

\section{\label{sec:level1}II. Experiment}

\begin{figure}
\includegraphics[width=8.5cm,height=13cm]{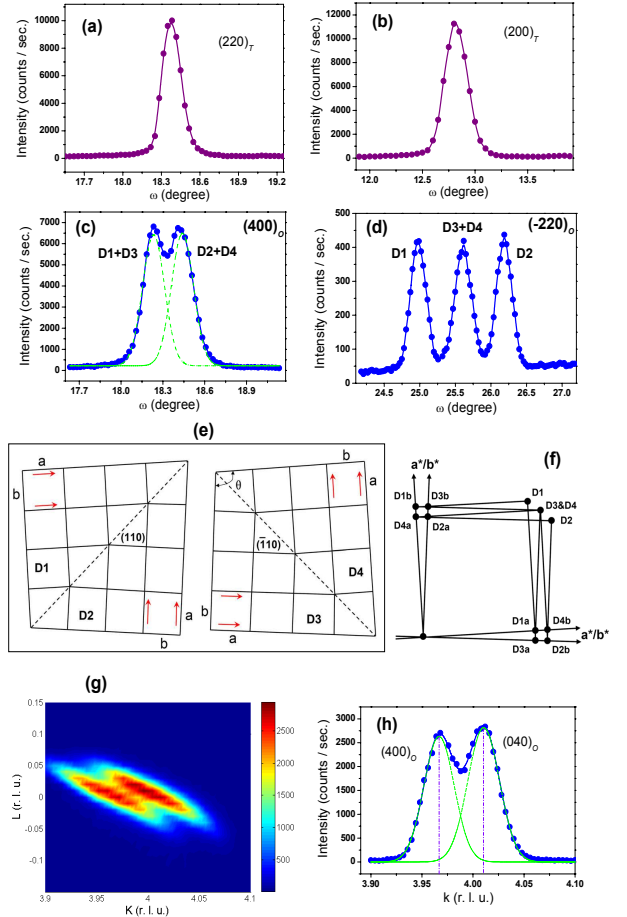}
\caption{\label{fig:epsart} (Color online) (a)(b) Omega scans of
tetragonal (220)$_T$ and (400)$_T$ nuclear reflection at 300 K,
respectively. (C) Omega scan of orthorhombic (400)$_O$ nuclear
reflection at 2.5 K. The splitting of the reflection indicates the
existence of twining. (d) Triple splitting of the rocking curve of
orthorhombic ($\bar{2}$20)$_O$ reflection at 2.5 K, as measured with
the (\emph{hk}0) aligned nearly in the horizontal scattering plane
(e) Schematic presentation of the twinned orthorhombic lattice in
real space. Four domain patterns are marked as D1-D4. \emph{a} and
\emph{b} denote long and short axis of orthorhombic lattice. Red
arrows indicate the direction of the Fe magnetic moment. (f)
Schematic presentation of the reciprocal space corresponding to the
twinned orthorhombic domains. (g) The contour map of orthorhombic
(400)$_O$ reflection at 2.5 K. (h) \emph{Q} scan of (400)$_O$
reflection.}
\end{figure}

EuFe$_2$As$_2$ single crystals were grown by the flux method. A
small amount of powdered single crystal was examined by means of
x-ray powder diffraction (XRD) analysis. The XRD pattern reveals a
single phase of EuFe$_2$As$_2$ in the tetragonal ThCr$_2$Si$_2$
structure with space group \emph{I4/mmm} at room temperature. The
samples have also been characterized via the measurements of heat
capacity, resistivity and magnetic susceptibility. Two prominent
phase transitions can be identified respectively at 190 and 19 K,
consistent to those previously reported \cite{Raffius,Ren2}. A 50 mg
single crystal with dimension  about 5~$\times$~5 $\times$~1~mm$^3$
was selected for neutron diffraction experiment, which was performed
on hot-neutron four-circle diffractometer HEIDI at FRM II, Garching
(Germany). A Cu (220) monochromator was selected to produce a
monochromatic neutron beam with the wavelength at 0.868 $\,
$\AA$^{}$. An Er filter was used to minimize the $\lambda$/2
contamination. Single crystal sample was mounted on a thin aluminium
holder inside a standard closed-cycle cryostat. The diffraction data
were collected using a $^3$He single detector at different
temperatures from 300 K down to 2.5 K. A fine collimation
($\thicksim$ 15$'$) in front of the sample and a narrow opening of
the detector slits were adopted to achieve a sufficient resolution,
in order to determine precisely the structural splitting due to
orthorhombic twinning and magnetic modulation wave vectors. To
ensure the inclusion of the contributions from all possible twinned
domains, the integrated intensities were collected with the setup
adopting a 60$'$ collimation and an angular opening of both
horizontal and vertical detector slits set at ~4.5 degree.
Furthermore, the integrated intensities for the reflections with
2$\theta$ $>$ 60$\textordmasculine$ and 2$\theta$ $<$
60$\textordmasculine$ were obtained respectively via the
$\theta$-2$\theta$ and the rocking-curve scans. The obtained data
used for the structural refinements were normalized by monitor
counts and corrected for the Lorentz factor. DATAP program is used
to carry out absorption correction by considering the size and shape
of crystal \cite{Coppens}. The absorption coefficient $\mu$ is
calculated to be 2.61 mm$^{-1}$ and the transmission factors are
deduced to be only in the range from 2.1$\%$ up to 14.2$\%$ due to
the extremely strong absorption. Determination of both the nuclear
and magnetic structures was performed by using the FULLPROF program
suit \cite{Carvajal}. The scale factor derived from the crystal
structure refinement was used to determine the magnitude of magnetic
moment from the magnetic reflections.

\section{\label{sec:level1}III. Results and Discussion}

\begin{figure}
\includegraphics[width=8.2cm,height=11.5cm]{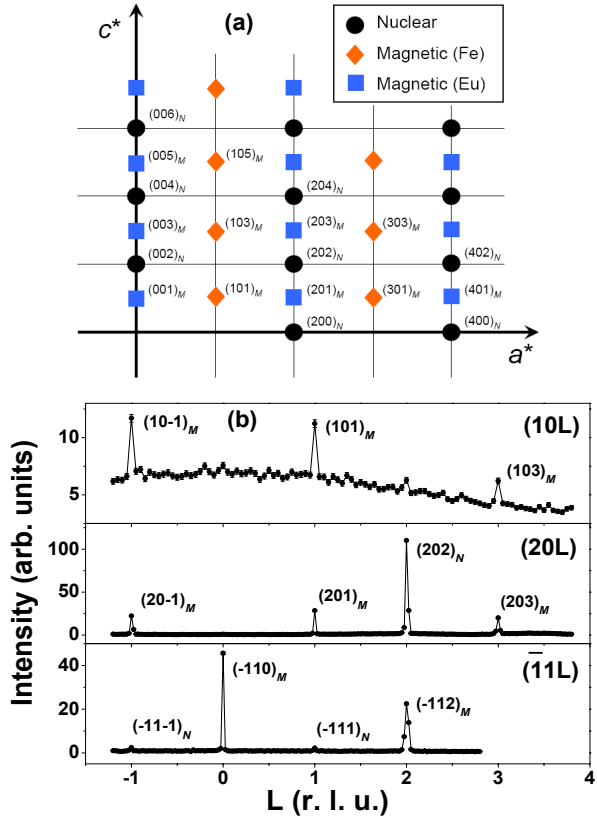}
\caption{\label{fig:epsart} (Color online) (a) The schematic diagram
of (\emph{h 0 l}) plane in the first quadrant of reciprocal space.
The circular, rhombic and square symbols represent the nuclear
reflection as well as the magnetic reflection attributed to Fe and
Eu magnetic sublattices. (b) Long \emph{l} scan on (01\emph{l}),
(02\emph{l}) and ($\bar{1}$1\emph{l}) reflections. }
\end{figure}

First of all, the crystal structure of EuFe$_2$As$_2$ is described
within the framework of tetragonal symmetry at 300 K. The $\omega$
scans of selected nuclear (220)$_T$ and (200)$_T$ reflections with
mosaic width of $\sim$0.22$\textordmasculine$ are shown in Fig. 1
(a) and (b), which indicate the good quality and homogeneity of the
single crystal. Upon cooling down, a splitting is observed for
orthorhombic (400)$_O$ and (040)$_O$ reflections (Fig. 1(c)). Note
that those two reflections are corresponding to the (220)$_T$ in
tetragonal setting according to the Tetragonal-Orthorhombic symmetry
relation. The observed splitting of (400)$_O$ is the indication of
T-O structural transition and accompanied twinning configuration due
to the interchange of the orthorhombic \emph{a} and \emph{b} axes.
It is known that twinning in orthorhombic structure will result in
four different domain patterns \cite{McIntyre, Tanatar}, as
illustrated in Fig. 1 (e). Two of domains shared the same (110)
plane and formed the domain pairs (D1 and D2), while another two
shared the ($\bar{1}$10) plane (D3 and D4). In principle, it is
possible to observe single peak, two or three or four peaks
depending on the selected reflections and the resolution of the
instrument. In Fig. 1(c), the left and right peaks can be assigned
to the contributions from the domains (D1+D3) and (D2+D4)
respectively. Note that the $\omega$-scan is performed with open
detector slits. Two Gaussian peaks were used to fit the (400)$_O$
and (040)$_O$ reflections and the domain population ratio is
estimated to be around 1:1 for the (\emph{h}00) and (0\emph{k}0)
twins, \emph{i.e.} D1+D3 $\approx$ D2+D4. The $\omega$-scan of
($\bar{2}$20) is examined afterward with the (\emph{hk}0) aligned in
the horizontal scattering plane to obtain more detailed information
about domain population (Fig. 1(d)). The occurrence of twinning and
T-O structural phase transition can be confirmed from the clear
presence of triple splitting of ($\bar{2}$20) nuclear reflection.
Usually, the reflections with \emph{h} = \emph{k}
$\textdoublebarslash$ 0 are triply split in twinned orthorhombic
lattice and the peak in the center is attributed to pairs that share
the same ($\bar{1}$10) or (110) plane, while the peak at left and
right sides corresponds to the rest two domains. In Fig. 1(d), it
can be seen that all three peaks showed almost equal intensity. This
strongly indicates that the domain population exists the following
relationship: D1 $\approx$ D3+D4 $\approx$ D2. Hence the domain
population for all those four different domain patterns can be
determined roughly as 2:2:1:1. In order to investigate the
distribution of nuclear reflection in reciprocal space and determine
the lattice parameter accurately, two dimensional plot of (400)$_O$
reflection is shown in Fig. 1 (g). The splitting of (400)$_O$ can
also be clearly seen. Totally 280 nuclear reflections were collected
for nuclear structure refinement within the \emph{Fmmm} space group.
Several strong reflections were excluded from the refinement because
of the significant extinction. All atoms were refined with the
isotropic temperature factor. The refinement results of crystal
structure are listed in Table 1. The lattice parameters are deduced
to be \emph{a} = 5.537(2) $\, $\AA$^{}$, \emph{b} = 5.505(2) $\,
$\AA$^{}$ and \emph{c} = 12.057(2) $\, $\AA$^{}$ at 2.5 K, which are
in good agreement with a previous report \cite{Tegel}.

\begin{figure}
\includegraphics[width=8.2cm,height=10cm]{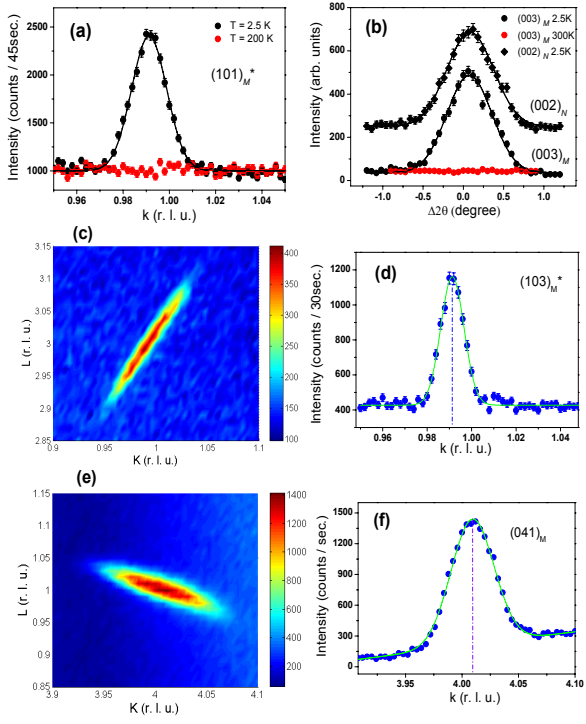}
\caption{\label{fig:epsart} (Color online) (a) The comparison of the
\emph{Q} scan of (101)$_M$ magnetic reflection at 2.5 and 200 K. The
(101)$_M$ reflection is observed in \emph{k} scan because of the
existence of twining. (b) The $\theta$-2$\theta$ scan of (003)$_N$
nuclear and (003)$_M$ magnetic reflections at 2.5 K, the same scan
of (003)$_M$ magnetic reflection at 300 K is also plotted for
comparison. (c) The contour map shows the Q dependence of the
(103)$_M$ magnetic reflection. (d) \emph{Q} scan of (103)$_M$
magnetic reflection. The (103)$_M$ reflection is observed in
\emph{k} scan because of the existence of twining. (e) The contour
map of (041)$_M$ magnetic reflection indicates the contribution of
the magnetic reflection of Eu magnetic sublattice. (f) \emph{Q} scan
of (041)$_M$ magnetic reflection. }
\end{figure}

To clarify the magnetic structure of EuFe$_2$As$_2$ at low
temperature, the sample was cooled to 2.5 K, which is well below the
reported Fe$^{2+}$ and Eu$^{2+}$ magnetic ordering temperatures.
Considering the existence of the twined (\emph{h}00) and
(0\emph{k}0) domains, extensive search of magnetic reflections was
performed in the \emph{a*}-\emph{c*} reciprocal space as
schematically illustrated in Fig. 2(a). Additional search was also
performed in the (\emph{hhl}) reciprocal plane. Fig. 2(b) shows
three typical long \emph{l} scans in the reciprocal space where in
addition to the expected nuclear reflections, two sets of magnetic
superstructure reflections can be clearly identified with two
magnetic propagation wave vectors (1,0,1) and (0,0,1) respectively.
As an example, \emph{Q} scan of (101)$_M$ magnetic reflection is
plotted in Fig. 3(a) and the same scan at 200 K is also plotted
together for comparison. In Fig. 3(b), the $\theta$-2$\theta$ scan
of nuclear (002)$_N$ and magnetic (003)$_M$ reflections show the
same peak center, which indicates that the magnetic structure is
commensurate in nature. The contour map of (103)$_M$ and (401)$_M$
reflections fully illustrated the intensity distribution as shown in
Fig. 3(c) and Fig. 3(e). As already discussed, the contour map of
(400)$_O$ nuclear reflection (Fig. 1(g)) clearly shows two
reflections attributed to the (\emph{h}00) and (0\emph{k}0) twins.
Two peak centers with \emph{k} = 3.967 and 4.01 can be obtained by
fitting the \emph{Q} scan of (400)$_O$ reflection(Fig. 1(h)). In
Fig. 3.(d), the \emph{Q} scan of (103)$_M$ reflection can be fitted
by a single Gaussian function with \emph{k} = 0.991. This strongly
indicates that the (\emph{h0l}) type reflections (with \emph{h} and
\emph{l} equal to odd number) are associated with the (\emph{h}00)
domain and they can thus be described with the propagation wave
vector \textbf{k} = (1,0,1). This wave vector is exactly the same as
observed in other '122' pnictides, such as BaFe$_2$As$_2$ \cite{Su1}
and CaFe$_2$As$_2$ \cite{Goldman}, which is related to the AFM order
of Fe$^{2+}$ moments. The magnetic structure refinement was then
carried out to determine the magnitude and direction of Fe$^{2+}$
moment. The magnetic structure with Fe saturation moment of 0.98(8)
$\mu$$_B$ aligned along the long \emph{a} axis is deduced. Note that
the origin of AFM order in FeAs-based pnictides is still a matter of
controversy. It is argued that the AFM order of Fe$^{2+}$ may arise
from the SDW order due to the Fermi surface instability under the
itinerant model \cite{Mazin}. While other evidences support that the
AFM order has a local moment origin as in Mott insulator
\cite{Si,Fang}, such as the parent compound of high \emph{T}$_c$
cuprates. Recent spin wave excitation study on CaFe$_2$As$_2$
suggests that the magnetism of iron arsenide might be resulted from
a complicated mixture of localized and itinerant properties and it
should be understood by considering both the localized and itinerant
electrons \cite{Zhao2}.

\begin{figure}
\includegraphics[width=8.2cm,height=10cm]{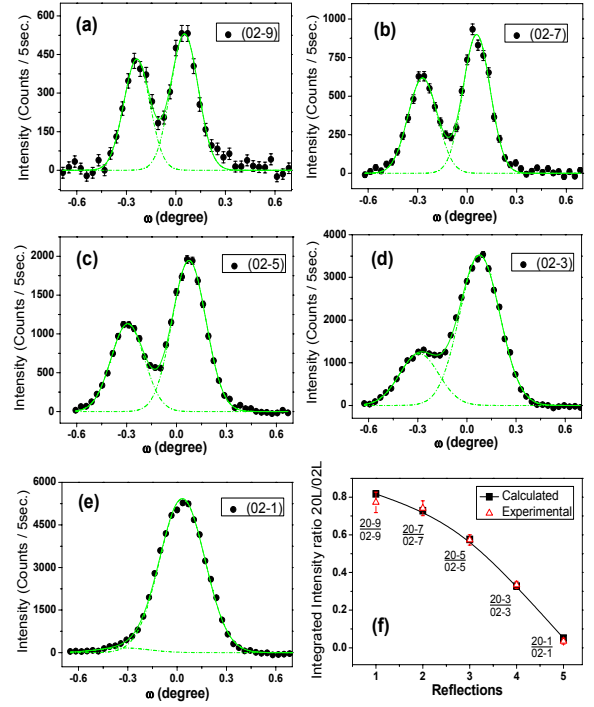}
\caption{\label{fig:epsart} (Color online) (a)-(e) Omega scans of
series of (02\emph{l}) (with \emph{l} = odd) reflections at 2.5 K.
The integrated intensities of (20\emph{l}) and (02\emph{l}) can be
obtained by fitting the curves with two Gaussian functions. (f) The
ratio between (20\emph{l}) and (02\emph{l}) reflections shows good
agreement with the calculated value.}
\end{figure}

\begin{table}
\caption{\label{tab:table1} Refined results of the crystal and
magnetic structures for EuFe$_2$As$_2$ at 2.5 K (space group:
\emph{Fmmm}, \emph{Z} = 4).}
\begin{ruledtabular}
\begin{tabular}{llldc}
Atom/site&\emph{x}&\emph{y}&\emph{z}&\emph{B}($\,$\AA$^2$)\\
\hline
Eu (4\emph{a})& 0 & 0& 0&0.81(3)\\
\quad \textbf{k}, \emph{M$_a$}($\mu$$_B$)& (0,0,1), 6.8(3)& & \\
Fe (8\emph{f})& 0.25 & 0.25 & 0.25 & 0.26(3)\\
\quad \textbf{k}, \emph{M$_a$}($\mu$$_B$)& (1,0,1), 0.98(8)& & \\
As (8\emph{i})& 0 & 0 & 0.363(5)& 0.25(3)\\
\hline
\multicolumn{5}{l}{\emph{a}, \emph{b}, \emph{c} ($\,$\AA): \quad 5.537(2), 5.505(2), 12.057(2)}\\
\multicolumn{5}{l}{Number of reflections (Nuclear):  \quad 280}\\
\multicolumn{5}{l}{\emph{RF}$^2$, \emph{RF}$^{2W}$, \emph{RF}($\%$), $\chi$$^2$ Nuclear: \quad 9.34, 9.67, 6.22, 7.1}\\
\multicolumn{5}{l}{Number of reflections (Magnetic): \quad 228}\\
\multicolumn{5}{l}{\emph{RF}$^2$, \emph{RF}$^{2W}$, \emph{RF}($\%$), $\chi$$^2$ Magnetic: \quad  9.42, 7.68, 6.53, 5.7}\\
\end{tabular}
\end{ruledtabular}
\end{table}

\begin{figure}
\includegraphics[width=7.5cm,height=5.3cm]{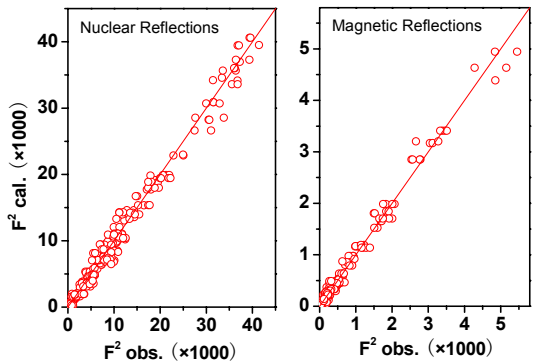}
\caption{\label{fig:epsart} (Color online) Integrated intensities of
the nuclear and magnetic Bragg reflections collected at 2.5 K are
plotted against the calculated values. See text for details of the
crystal and magnetic structure models.}
\end{figure}

Consequently, the magnetic reflections with a propagation wave
vector \textbf{k} = (0,0,1) (with \emph{h} even and \emph{l} odd)
are due to the long range order of the localized Eu$^{2+}$ moments.
However, the moment direction of Eu$^{2+}$ moments is still
indeterminate. Symmetry analysis based on the representation theory
indicates that the magnetic representation $\Gamma$ for magnetic
Eu$^{2+}$ on 4\emph{a} site is decomposed into three one dimensional
irreducible representations: $\Gamma$$_1$, $\Gamma$$_2$ and
$\Gamma$$_3$. The Eu$^{2+}$ moments are aligned in the \emph{c},
\emph{a} or \emph{b} direction according to those three
representations. The observation of nonzero intensity of
(00\emph{l})(with \emph{l} = odd) reflections clearly exclude the
representation $\Gamma$$_1$. The $\omega$ scans on several
(\emph{hk}0)$_M$ (with both \emph{h} and \emph{k} = odd) reflections
also exhibit considerable intensity. Thus, the moment of Eu$^{2+}$
is expected to be aligned either along the \emph{a} or \emph{b}
direction in the \emph{ab} plane. The \emph{Q} scan on (041)$_M$
reflection (Fig. 3(f)) giving a peak position of \emph{k} = 4.01,
which is exactly equal to the larger \emph{k} value of (400)$_O$
nuclear reflection. Therefore, the moment direction of Eu$^{2+}$ can
be determined as along the \emph{a} direction since the intensity
ratio between (041)$_M$ and (401)$_M$ magnetic reflections
approximate equals to 73:1 for this arrangement. The structure mode
is confirmed further by $\omega$ scan of series of (02\emph{l})(with
\emph{l} = odd) reflections as shown in Fig. 4. Similar to some
nuclear reflections, both (20\emph{l}) and (02\emph{l}) magnetic
reflections was detected due to the twinning configuration. However,
the intensity ratio between (20\emph{l}) and (02\emph{l}) changes
gradually with the change of the angle between the scattering plane
and the \emph{c} axis. The calculated intensity ratio of
(20\emph{l})/(02\emph{l}) for different \emph{l} are plotted in Fig.
4(f) and it agrees well with the observed values which derived from
the $\omega$ scans directly. By taking into account of twinning
components properly, the refinement on Eu$^{2+}$ magnetic sublattice
was carried out with the aforementioned magnetic structure model.
The calculated structure factors are plotted against those observed
and shown in Fig. 5. The reliable agreement factors confirms the
proposed magnetic structural model eventually, \emph{i.e.} the
Eu$^{2+}$ moment aligns along \emph{a} direction with the wave
vector \textbf{k} = (0,0,1) and magnitude of 6.8(3) $\mu$$_B$. Thus
the magnetic structure of EuFe$_2$As$_2$ is unambiguously determined
as illustrated in Fig. 6. The moment direction of Eu$^{2+}$ is also
consistent with our resonant x-ray scattering (RXS) measurement
\cite{Javier}.

\begin{figure}
\includegraphics[width=6.2cm,height=7cm]{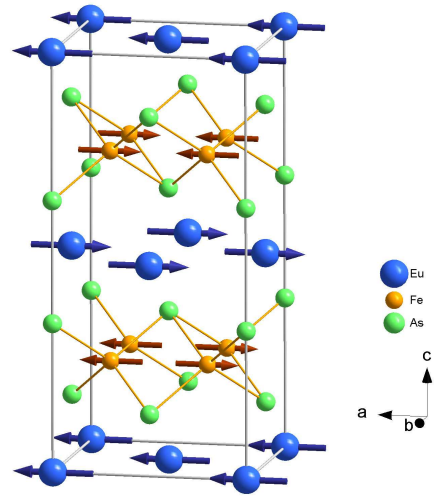}
\caption{\label{fig:epsart} (Color online) Illustration of the
magnetic structures of EuFe$_2$As$_2$ at 2.5 K. The Fe moments align
along \emph{a} direction and order antiferromagnetically in both
\emph{a} and \emph{c} directions. The Eu moments align along
\emph{a} direction and order antiferromagnetically in \emph{c}
direction only. The gray line outlines the orthorhombic unit cell.}
\end{figure}

Fig. 7(a) shows the temperature dependence of the (112)$_M$ and
(003)$_M$ magnetic reflections attributed to the ordering of
Eu$^{2+}$ moments. The onset temperature of Eu$^{2+}$ magnetic order
is deduced to be 19 K, which is in good agreement with previous
report on electronic and magnetic measurements \cite {Ren2,Raffius}.
The magnetic ordering temperature of Fe$^{2+}$ moment is estimated
to be 190 K based on the temperature dependence of the (101)$_M$ and
(103)$_M$ magnetic reflections (see Fig. 7(b)). The T-O structural
phase transition also takes place at 190 K as revealed by the sharp
change of full width at half maximum (FWHM) in (040)$_O$ nuclear
reflection. First principle calculations suggest that the nearest
and next nearest neighbor superexchange interactions between Fe ions
lead to a frustrated magnetic ground state in pnictides with
parallel and antiparallel arrangement of Fe spins in FeAs layer
\cite{Yildirim}. Usually, the magnetic frustration can be lifted by
a structural distortion from low symmetry to high symmetry phase.
This may be the origin of  the strong coupling between the
structural and magnetic phase transitions observed in EuFe$_2$As$_2$
and other iron pnictides \cite{Huang,Su1,Goldman}.

\begin{figure}
\includegraphics[width=8.5cm,height=5.3cm]{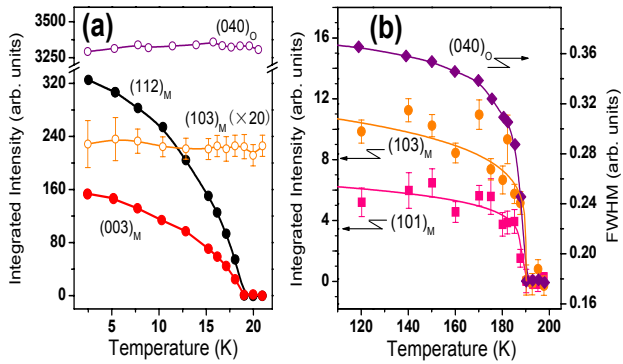}
\caption{\label{fig:epsart} (Color online) (a) Temperature
dependence of integrated intensity of (040)$_O$ nuclear reflection
as well as the (103)$_M$ , (112)$_M$ and (003) magnetic reflections
below 22 K. (b) Temperature dependence of integrated intensity of
(103)$_M$ and (101)$_M$ reflections; temperature dependence of FWHM
of (040)$_o$ reflection. }
\end{figure}

Due to the localized nature of Eu 4\emph{f} state, the AFM coupling
of Eu$^{2+}$ moments would be described by the indirect exchange,
\emph{e.g.} the Ruderman-Kittel-Kasuya-Yosida (RKKY) interaction as
suggested by Ren \emph{et al} \cite{Ren3}. Besides of the Eu-Eu and
Fe-Fe interactions, the strength of the interaction between the Eu
and Fe magnetic sublattices is also an interesting issue. Similar to
CaFe$_2$As$_2$, the SDW transition in EuFe$_2$As$_2$ can also be
suppressed continuously by applying the pressure due to the
weakening of nearest Fe-Fe exchange coupling \cite{Miclea}. Whereas
the AFM ordering temperature of Eu sublattice does not change
significantly even the compound exhibits the possible reentrant
superconducting state. This may suggest the weak interaction between
the Eu and Fe magnetic sublattices, which is supported by the full
potential electronic structure calculation \cite{Jeevan2}. In
present neutron work, we also did not observe any detectable change
in Fe$^{2+}$ magnetic moment when temperature passing through the
Eu$^{2+}$ magnetic ordering temperature (Fig. 7(a)). The coupling
between Fe$^{2+}$ and Eu$^{2+}$ moments in the ordered state
varnishes entirely within a mean field model due to geometrical
frustration. Parallel alignment of the Fe$^{2+}$ and Eu$^{2+}$
moments might be resulted from thermal or ground state fluctuations,
as suggested in an order by disorder scheme \cite{Villain, Reimers}.
Those results are in contrast to some '1111' compounds, such as
PrFeAsO, in which the interplay between Fe and Pr ordering moments
might drive the negative thermal expansion \cite{Kimber}.

\section{\label{sec:level1}IV. Conclusion}

Single crystal neutron diffraction experiment using a hot neutron
source was performed to investigate the crystal and magnetic
structure of EuFe$_2$As$_2$. With decreasing temperature, the
antiferromagnetic order of Fe$^{2+}$ moments set in at 190 K with
the propagation vector \textbf{k} = (1,0,1). Similar to
BaFe$_2$As$_2$ and CaFe$_2$As$_2$, the tetragonal to orthorhombic
structural transition occurs simultaneously with the AFM order,
which indicates the strong coupling between the lattice and Fe
magnetic degree of freedom. Below 19 K, the Eu$^{2+}$ moments order
antiferromagneticly with the propagation vector \textbf{k} = (1,0,1)
and are aligned along the \emph{a} axis. our studies also suggest a
weak coupling between the Fe$^{2+}$ and Eu$^{2+}$ magnetic
sublattice.

\section{\label{sec:level1}ACKNOWLEDGMENT}

We would like to thank G. McIntyre from the Institut Laue-Langevin
(France) for his kind help in the data analysis.

\appendix


\begin{thebibliography}{10}

\bibitem{Kamihara}
Y. Kamihara, T. Watanabe, M. Hirano, and H. Hosono, J. Am. Chem.
Soc. \textbf{130}, 3296 (2008).

\bibitem{Takahashi}
H. Takahashi, K. Igawa, K.Arii, Y. Kamihara, M. Hirano, and H.
Hosono, Nature (London) \textbf{453}, 376 (2008).


\bibitem{Chen1}
X. H. Chen, T. Wu, G. Wu, R. H. Liu, H. Chen, and D. F. Fang, Nature
(London) \textbf{453}, 761 (2008).


\bibitem{Matsuishi1}
S. Matsuishi,Y. Inoue, T. Nomura, H. Yanagi, M. Hirano, and H.
Hosono, J. Am. Chem. Soc. \textbf{130}, 14428 (2008).

\bibitem{Rotter1}
M. Rotter, M. Tegel, and D. Johrendt, Phys. Rev. Lett. \textbf{101},
107006 (2008).



\bibitem{Wen}
H. H. Wen, G. Mu, L. Fang, H. Yang, and X. Zhu, Europhys. Lett.
\textbf{82}, 17009 (2008).

\bibitem{Ren}
Z.-A. Ren, W. Lu, J. Yang, W. Yi, X.-L. Shen, Z.-C. Li, G.-C. Che,
X.-L. Dong, L.-L. Sun, F. Zhou, and Z.-X. Zhao, Chin. Phys. Lett.
\textbf{25}, 2215 (2008).


\bibitem{Matsuishi2}
S. Matsuishi, Y. Inoue, T. Nomura, M. Hirano, and H. Hosono, J.
Phys. Soc. Jpn. \textbf{77}, 113709 (2008).



\bibitem{Rotter2}
M. Rotter, M. Tegel, and D. Johrendt, Phys. Rev. Lett. \textbf{101},
107006 (2008).


\bibitem{Cheng}
Peng Cheng, Bing Shen, Gang Mu, Xiyu Zhu, Fei Han, Bin Zeng, and
Hai-Hu Wen, arXiv:0812.1192 (2008).



\bibitem{Cruz}
C. de la Cruz, Q. Huang, J. W. Lynn, J. Li, W. Ratcliff II, J. L.
Zarestky, H. A. Mook, G. F. Chen, J. L. Luo, N. L. Wang, and P. C.
Dai, Nature (London) \textbf{453}, 899 (2008).


\bibitem{Huang}
Q. Huang, Y. Qiu, W. Bao, J.W. Lynn, M. A. Green, Y. Chen, T. Wu, G.
Wu, and X. H. Chen, Phys. Rev. Lett. \textbf{101}, 257003 (2008).

\bibitem{Su1}
Y. Su, P. Link, A. Schneidewind, Th. Wolf, Y. Xiao, R. Mittal, M.
Rotter, D. Johrendt, Th. Brueckel, and M. Loewenhaupt, Phys. Rev. B
\textbf{79}, 064504 (2009).


\bibitem{Goldman}
A. I. Goldman, D. N. Argyriou, B. Ouladdiaf, T. Chatterji, A.
Kreyssig, S. Nandi, N. Ni, S.L. Bud'ko, P. C. Canfield, and R. J.
McQueeney, Phys. Rev. B \textbf{78}, 100506(R) (2008).


\bibitem{Xiao}
Y. Xiao, Y. Su, R. Mittal, T. Chatterji, T. Hansen, C.M.N. Kumar, S.
Matsuishi, H. Hosono, and Th. Brueckel, Phys. Rev. B \textbf{79},
060504(R) (2009).


\bibitem{Zhao1}
J. Zhao, Q. Huang, C. de la Cruz, S. Li, J. W. Lynn, Y. Chen, M. A.
Green, G. F. Chen, G. Li, Z. Li, J. L. Luo, N. L. Wang, and P. Dai,
Nature Materials \textbf{7}, 953 (2008).


\bibitem{Bednorz}
J. G. Bednorz, and K. A. Mueller, Z. Phys. B \textbf{64}, 189
(1986).


\bibitem{Ranjan1}
R. Mittal, Y. Su, S. Rols, T. Chatterji, S. L. Chaplot, H. Schober,
M. Rotter, D. Johrendt, and Th. Brueckel, Phys. Rev. B \textbf{78},
104514 (2008).


\bibitem{Ranjan2}
R. Mittal, Y. Su, S. Rols, M. Tegel, S. L. Chaplot, H. Schober, T.
Chatterji, D. Johrendt, and Th. Brueckel, Phys. Rev. B \textbf{78},
224518 (2008).

\bibitem{Christianson}
A. D. Christianson, E. A. Goremychkin, R. Osborn, S. Rosenkranz, M.
D. Lumsden, C. D. Malliakas, I. S. Todorov, H. Claus, D. Y. Chung,
M. G. Kanatzidis, R. I. Bewley and T. Guidi, Nature \textbf{456},
930 (2008).

\bibitem{Lumsden}
M. D. Lumsden, A. D. Christianson, D. Parshall, M. B. Stone, S. E.
Nagler, G.J. MacDougall, H. A. Mook, K. Lokshin, T. Egami, D. L.
Abernathy, E. A. Goremychkin, R. Osborn, M. A. McGuire, A. S. Sefat,
R. Jin, B. C. Sales, and D. Mandrus, Phys. Rev. Lett. \textbf{102},
107005 (2009).

\bibitem{Chi}
Songxue Chi, Astrid Schneidewind, Jun Zhao, Leland W. Harriger,
Linjun Li, Yongkang Luo, Guanghan Cao, Zhu'an Xu, Micheal
Loewenhaupt, Jiangping Hu, and Pengcheng Dai, Phys. Rev. Lett.
\textbf{102}, 107006 (2009).

\bibitem{Su2}
Y. Su \emph{et al.}, (in preparation) (2009).

\bibitem{Ren2}
Zhi Ren, Zengwei Zhu, Shuai Jiang, Xiangfan Xu, Qian Tao, Cao Wang,
Chunmu Feng, Guanghan Cao, and Zhu'an Xu, Phys. Rev. B \textbf{78},
052501 (2008).

\bibitem{Raffius}
H. Raffius, M. Moersen, B. D. Mosel, W. Mueller-Warmuth, W.
Jeitschko, L. Terbuechte, and T. Vomhof, J. Phys. Chem. Solids
\textbf{54}, 135 (1993).

\bibitem{Jiang}
Shuai Jiang, Yongkang Luo, Zhi Ren, Zengwei Zhu, Cao Wang, Xiangfan
Xu, Qian Tao, Guanghan Cao, and Zhu'an Xu, New J. Phys. \textbf{11},
025007 (2009).

\bibitem{Wu}
T. Wu, G. Wu, H. Chen, Y. L. Xie, R. H. Liu, X. F. Wang, and X. H.
Chen, arXiv:0808.2247.


\bibitem{Jeevan1}
H. S. Jeevan, Z. Hossain, Deepa Kasinathan, Helge Rosner, C. Geibel,
and P. Gegenwart, Phys. Rev. B \textbf{78}, 092406 (2008).

\bibitem{Qi}
Yanpeng Qi, Zhaoshun Gao, Lei Wang, Dongliang Wang, Xianping Zhang,
and Yanwei Ma, New Journal of Physics \textbf{10} 123003 (2008).


\bibitem{Ren3}
Zhi Ren, Xiao Lin, Qian Tao, Shuai Jiang, Zengwei Zhu, Cao Wang,
Guanghan Cao, and Zhu'an Xu, Phys. Rev. B \textbf{79}, 094426
(2009).



\bibitem{Coppens}
P. Coppens, L. Leiserowitz, and D. Rabinovich, Acta Crystallogr.
 \textbf{18}, 1035 (1965).


\bibitem{Carvajal}
J. L. Rodr¨ªguez-Carvajal, Physica B \textbf{192}, 55 (1993).


\bibitem{McIntyre}
G. J. McIntyre, A. Renault, and G. Collin, Phys. Rev. B \textbf{37},
5148 (1988).

\bibitem{Tanatar}
M. A. Tanatar, A. Kreyssig, S. Nandi, N. Ni, S. L. Bud'ko, P. C.
Canfield, A. I. Goldman, and R. Prozorov, Phys. Rev. B \textbf{79},
180508(R) (2009).


\bibitem{Tegel}
Marcus Tegel, Marianne Rotter, Veronika Weiss, Falko M. Schappacher,
Rainer Poettgen, and Dirk Johrendt, J. Phys.: Condens. Matter
\textbf{20}, 452201 (2008).


\bibitem{Mazin}
I. I. Mazin, D. J. Singh, M. D. Johannes, and M. H. Du, Phys. Rev.
Lett. \textbf{101}, 057003 (2008).



\bibitem{Si}
Qimiao Si and Elihu Abrahams, Phys. Rev. Lett. \textbf{101}, 076401
(2008).

\bibitem{Fang}
Chen Fang, Hong Yao, Wei-Feng Tsai, JiangPing Hu, and Steven A.
Kivelson, Phys. Rev. B \textbf{77}, 224509 (2008).

\bibitem{Zhao2}
Jun Zhao, D. T. Adroja, Dao-Xin Yao, R. Bewley, Shiliang Li, X. F.
Wang, G. Wu, X. H. Chen, Jiangping Hu, and Pengcheng Dai,
arXiv:0903.2686.


\bibitem{Javier}
Javier Herrero-Martin, Valerio Scagnoli, Claudio Mazzoli, Yixi Su,
Ranjan Mittal, Y. Xiao, Th. Brueckel, Kumar Neeraj, S. K. Dhar, A.
Thamizhavel and Luigi Paolasini, arXiv:0906.1508 (2009).



\bibitem{Yildirim}
T. Yildirim, Phys. Rev. Lett. \textbf{101}, 057010 (2008).




\bibitem{Miclea}
C. F. Miclea, M. Nicklas, H. S. Jeevan, D. Kasinathan, Z. Hossain,
H. Rosner, P. Gegenwart, C. Geibel, and F. Steglich,
arXiv:0808.2026.


\bibitem{Jeevan2}
H. S. Jeevan, Z. Hossain, Deepa Kasinathan, H. Rosner, C. Geibel,
and P. Gegenwart, Phys. Rev. B \textbf{78}, 052502 (2008).


\bibitem{Villain}
J. Villain, R. Bidaux, J.P. Carton and R. Conte, J. Physique
\textbf{41}, 1263 (1980).


\bibitem{Reimers}
Jan N. Reimers, and A. J. Berlinsky, Phys. Rev. B \textbf{48}, 9539
(1993).

\bibitem{Kimber}
S. A. J. Kimber, D. N. Argyriou, F. Yokaichiya, K. Habicht, S.
Gerischer, T. Hansen, T. Chatterji, R. Klingeler, C. Hess, G. Behr,
A. Kondrat, and B. Buechner, Phys. Rev. B \textbf{78}, 140503
(2008).


\end{thebibliography}
\end{document}